\documentclass[fleqn,usenatbib]{mnras}
\pdfoutput=1

\usepackage{newtxtext,newtxmath}

\usepackage[T1]{fontenc}
\usepackage{ae,aecompl}

\usepackage{graphicx}	
\usepackage{amsmath}	
\usepackage{amssymb}	
\usepackage{subfigure}

\title[Accretion in Thin MADs]{Angular Momentum Transport in Thin Magnetically Arrested Disks}

\author[M. D. Marshall et al.]{
Megan D. Marshall,$^{1}$\thanks{E-mail: meganmarshall20@gmail.com}
Mark J. Avara,$^{2}$
and Jonathan C. McKinney$^{1,3}$
\\
$^{1}$University of Maryland at College Park, Dept. of Physics, 3260 Physical Sciences Complex, College Park, MD 20742, USA\\
$^{2}$Rochester Institute of Technology, Center for Computational Relativity and Gravitation, 170 Lomb Memorial Dr, Rochester, NY 14623\\
$^{3}$Joint Space-Science Institute,1113 Physical Sciences Complex, College Park, MD 27042, USA
}

\date{Accepted XXX. Received YYY; in original form ZZZ}

\pubyear{2016}

\begin{document}
\label{firstpage}
\pagerange{\pageref{firstpage}--\pageref{lastpage}}
\maketitle

\begin{abstract}
In accretion disks with large-scale ordered magnetic fields, the magnetorotational instability (MRI) is marginally suppressed, so other processes may drive angular momentum transport leading to accretion. Accretion could then be driven by large-scale magnetic fields via magnetic braking, and large-scale magnetic flux can build-up onto the black hole and within the disk leading to a magnetically-arrested disk (MAD).  Such a MAD state is unstable to the magnetic Rayleigh-Taylor (RT) instability, which itself leads to vigorous turbulence and the emergence of low-density highly-magnetized bubbles.  This instability was studied in a thin (ratio of half-height H to radius R, $H/R \approx 0.1$) MAD simulation, where it has a more dramatic effect on the dynamics of the disk than for thicker disks. Large amounts of flux are pushed off the black hole into the disk, leading to temporary decreases in stress, then this flux is reprocessed as the stress increases again. Throughout this process, we find that the dominant component of the stress is due to turbulent magnetic fields, despite the suppression of the axisymmetric MRI and the dominant presence of large-scale magnetic fields.  This suggests that the magnetic RT instability plays a significant role in driving angular momentum transport in MADs.
\end{abstract}

\begin{keywords}
accretion -- accretion discs -- black hole physics -- gravitation
\end{keywords}



\section{Introduction}

Accretion disks are a central focus of interest in high energy astrophysics. From observations of AGN and similar objects, it is known that matter is flowing onto the central black hole (BH) of the system due to the release of intense radiation and emergence of powerful jets. \citet{1973A&A....24..337S} treated angular momentum transport in accretion disks as parameterized by an effective viscosity that was likely magnetic in origin. They described the stress using the so-called $\alpha$ viscosity prescription.

A critical breakthrough was when \citet{1991ApJ...376..214B, 1998RvMP...70....1B} found that weak magnetic fields are unstable to differential rotation and this can drive angular momentum radially outward, allowing gas to move radially inward.  This instability leads to an effective viscosity by ultimately leading to turbulence and a self-sustaining dynamo.  However, in disks with strong magnetic fields, suggested by \citet{2003PASJ...55L..69N}, the axisymmetric MRI is suppressed, meaning some new mechanism may be driving the angular momentum transport.

One possible source of stress leading to accretion onto compact objects is the effective viscosity generated by turbulence driven by the magnetic RT instability, as suggested in \citet{1976Natur.262..356E} and \citet{1976ApJ...207..914A}. Since then, it has been studied at magnetosphere-accretion disk boundaries near compact objects \citep{1992ApJ...386...83K, 1995ApJ...445..337L,1995MNRAS.275.1223S,  2004ApJ...601..414L,2007ApJ...671.1726S} as well as in more specific context of neutron star \citep{1983A&A...118..267W,1984A&A...135...66W,1985ApJ...299...85W,1990A&A...229..475S,2008ApJ...673L.171R,2008MNRAS.386..673K,2012MNRAS.421...63R,2016MNRAS.459.2354B} and BH environments \citep{2008ApJ...677..317I, 2012MNRAS.423.3083M,2012JPhCS.372a2040T,2016MNRAS.462..636A}. The magnetic barrier is unstable and leads to null points in the magnetic field, allowing magnetic interchange between magnetic flux and mass.  Once turbulence develops, a spectrum of modes develops.  The large-scale modes correspond to a large portion of magnetic flux being ripped off the black hole.  This magnetic flux pushes back into the disk, creating a low-density region, which we refer to as a bubble, in its wake. This process occurs many times during the thin (half-height H to radius R, $H/R \approx 0.1$) MADiHR simulation presented in \citet{2016MNRAS.462..636A}.  Both the magnetic RT-driven turbulence and the large-scale magnetic flux seem likely drivers of angular momentum transport.

We study the magnetic-RT-driven turbulence and the large-scale magnetic interchange events and how these affect the effective viscosity and accretion rate of the disk. In Section \ref{sec:Methods}, we describe the selection and visualization techniques developed as well as the stress calculations that were done. We discuss the results in Section~\ref{sec:results}, and provide conclusions in Section~\ref{sec:conclusion}.

\section{Methods}
\label{sec:Methods}
In this study, we use the MADiHR initially MAD thin disk ($H/R \approx 0.1$) simulation around a BH with dimensionless spin of $a/M = 0.5$, where $a$ is the BH spin and $M$ is the mass of the BH, presented in \citet{2016MNRAS.462..636A}. In the simulation, there are many times when the disk is disrupted by the magnetic flux coming off the BH due to the magnetic RT instability, the largest of these being the focus of this study.

\begin{figure}
\centering
\includegraphics[width=\columnwidth]{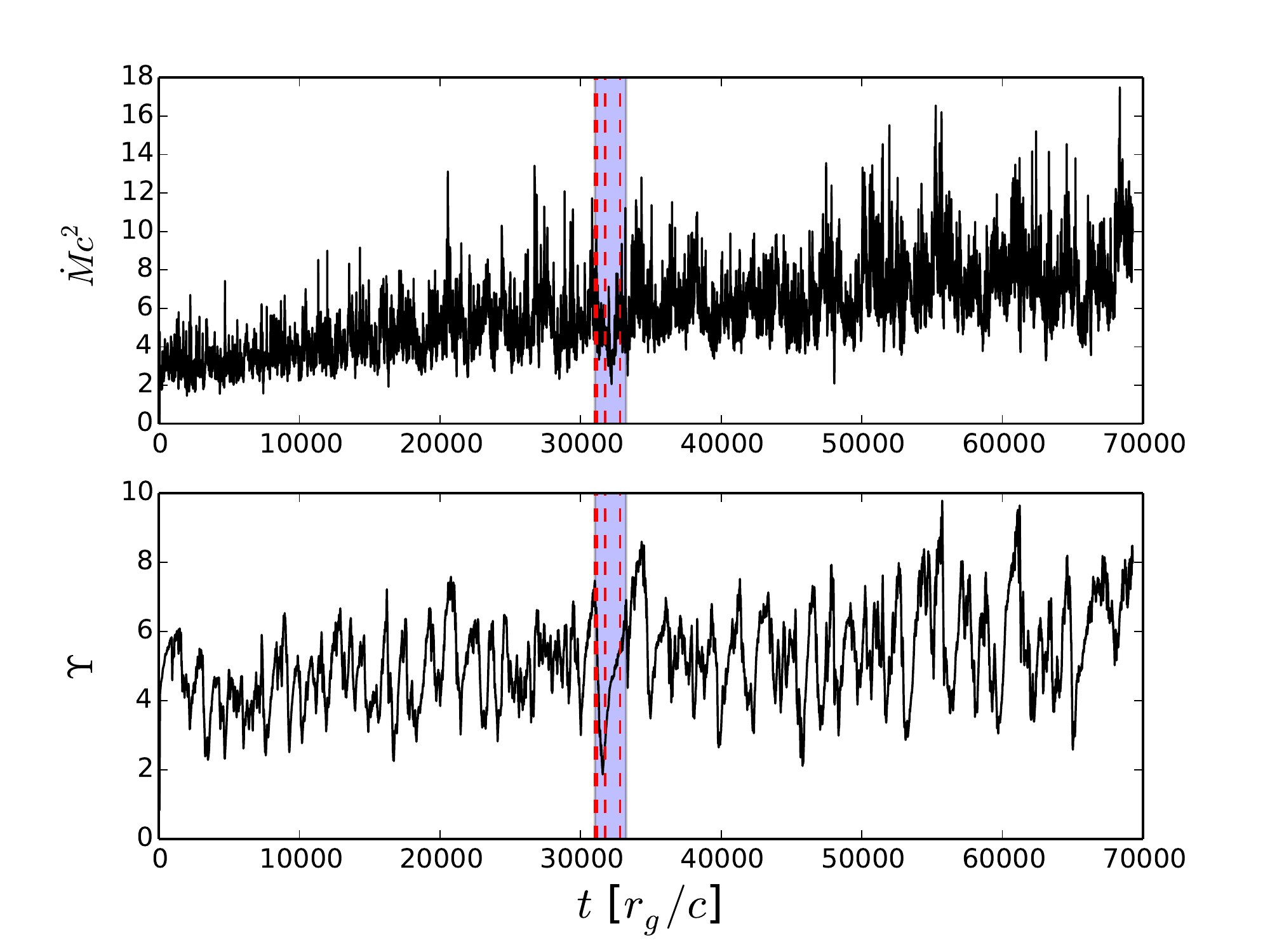}
\caption{Quantities used to define the lifetime of the magnetic RT bubble. The upper panel shows the mass accretion rate through the BH, while the bottom panel shows the normalized magnetic flux threading it. The time period being studied is highlighted in blue with red dashed lines marking the times displayed in Fig. \ref{fig:bubblesnaps}. The normalized flux provides a clearer picture of the dynamics around the BH, so it is used to determine the interesting time period. As magnetic Rayleigh-Taylor events occur, large amounts of magnetic flux move off the BH, but return to the BH as the bubble dissipates, so it is used to determine the lifespan of the RT bubble. For this project, we looked at the drop in magnetic flux that occurs between 31016 $r_{g}/c$ and 33240$ r_{g}/c$.}
\label{fig:upsilon}
\end{figure}

To study the effects the magnetic RT instability has on the accretion rate of the disk, we need to know the mass accretion rate $\dot{M}$,
\begin{equation}
\dot{M}=\left|\int\rho u^r dA_{\theta\phi} \right| ,
\end{equation}
where $\rho$ is the density, $u^r$ is the radial 4-velocity, and $dA_{\theta\phi}$ is the differential surface area, and  $\Upsilon_{H}$, the strength of the dimensionless magnetic flux on the horizon,
\begin{equation}
\Upsilon_{H}(r) \approx 0.7\frac{\int dA_{\theta\phi} 0.5|B^r|}{\sqrt{\langle\dot{M}_H\rangle}} \bigg\rvert_{r=r_{H}} ,
\end{equation}
with horizon radius $r_{H}$, radial magnetic field strength $B^r$ in Heaviside-Lorentz units and time averaged $\dot{M}$ on the horizon $\langle\dot{M}_H\rangle\approx 5.75$. $\Upsilon_{H}$ drops as large amounts of magnetic flux pushes into the disk as a result of the magnetic RT instability, shown in Fig. \ref{fig:upsilon}, so it can be used to define the lifespan of the RT bubble we studied, which is associated with the largest drop in $\Upsilon_{H}$. We define the time period of interest by finding the two maxima around this decrease in $\Upsilon_H$, showing the bubble's emerges at 31016 $r_{g}/c$ and dissipates at 33240 $r_{g}/c$.

After identifying the period of interest, we study the RT bubble using both visualizations (Section \ref{sec:vis}) and calculations (Section \ref{sec:stress}).

\subsection{Visualization}
\label{sec:vis} 
In this paper, we decide first to perform a qualitative analysis of the magnetic RT instability.  Detailed instability analysis are difficult (\citet{2004ApJ...601..414L}), and in the magnetic RT case are limited to very simple magnetic, density, and velocity profiles.  Like \citet{2004ApJ...606.1083H} and building on the density profiles of the equatorial plane of \citet{2008ApJ...677..317I}, we sought to use 3D renderings to understand the structure of the magnetic field.  We also generate a movie showing the evolution of the field. These renderings were made using vis5D+ \footnote{Freely available at: https://github.com/pseudotensor/Vis5dPlus .}.
Following prior work by \citet{2012MNRAS.423.3083M}, to generate the frames, the coordinate basis quantities are converted to an orthonormal basis using the full metric in Kerr-Schild coordinates. This orthonormal basis is then converted to spherical polar coordinates and then Cartesian coordinates.  We do not convert to Boyer-Lindquist time, and instead stick with Kerr-Schild time that is horizon-penetrating.  The data is also interpolated from the original grid to a Cartesian grid.  A resolution of $400^{3}$ grid cells covers the inner region from -40 $r_g$ to 40 $r_g$ to focus on the inner region of the disk close to the black hole.

To track the magnetic field evolution through the lifetime of the RT bubble, we select certain magnetic fieldlines in the low density region and follow them through the time period of interest. We initially select fieldline seedpoints in the RT bubble and disk midplane from the areas with high magnetic flux. This is done by creating an unnormalized probability distribution for the midplane. The function used was 
\begin{equation}
	P\left( r,\phi \right) = \begin{cases}
	\left|\frac{B_z a_{H}}{\sqrt{\langle\dot{M}\rangle} \langle \Upsilon_H \rangle}\right|, & \text{for } r\leq r_{H} \\
    \\
    \left|\frac{\frac{r}{r_H} B_z a_{H}}{\sqrt{\langle\dot{M}\rangle} \langle \Upsilon_H \rangle}\right|, & \text{for } r > r_{H}
	\end{cases},
    \label{eqn: prob}
\end{equation}
with vertical magnetic field strength $B_z$, $a_{H}$ is the surface area of the upper half of the horizon, and time-averaged $\langle \Upsilon_H \rangle \approx 5.0$ as reported in \citet{2016MNRAS.462..636A}. We developed this prescription based on $\Upsilon_H$, with a dimensionless factor added to the disk term to remove the radial dependence of the magnetic field. To normalize this probability for use in selecting seedpoints, points with $P\leq 10$ were set to 0 and $P_{max} \left( t \right) =5985.1$ is set to 1, and a linear function was created from those fixed values. We chose to use the maximum single cell value of $P\left( t \right)$ rather than an arbitrary cap to allow the probability function to most accurately represent the high flux regions.

After the initial points are chosen, they are propagated forward in time using the local fluid flow velocity. While this neglects magnetic diffusion processes, such as reconnection, it does a reasonable job of keeping the field tied to the fluid flow. To ensure the timestep used for each field line was small enough to accurately follow its motion, the local Keplarian period is compared to the time between data files. If the Keplarian period is much shorter than this time, an integer number of substeps is used to interpolate between the data files, with the seedpoints' new locations calculated using a velocity that is linearly interpolated between data files. After each step forward, the flux of the new position is checked using Equation \ref{eqn: prob}. If the new probability is 0, meaning the point is now outside the RT bubble, we drop that point and a new one is randomly chosen to replace it.

To visually represent the amount of flux in the disk at a given time, we vary the number of seedpoints displayed. As the flux in the disk grew, the number of seedpoints $N(t)$ at a given time t ranged from $N_{min}=15$ to $N_{max}=30$ as 
\begin{equation}
	N(t)=max \left( N_{max} \frac{\sum{P(t)}}{(\sum{P})_{max}},N_{min}\right) ,
\end{equation}
increasing as flux moved into the disk from the BH and decreasing as the flux was reprocessed by the disk. Finally, we fix a set of four fieldlines to $r=0.75r_H$ initially evenly spaced in $\phi$ over the upper hemisphere of the horizon to show the evolution of the field in the jet region during this time. In Fig. \ref{fig:bubblesnaps} below, we show the field structure at four characteristic times: before the bubble forms, the emergence of the bubble from the BH, the peak of the bubble's size, and the dissipation of the bubble.

\begin{figure*}
\centering
\subfigure[Quiescent]{\label{fig:a}\includegraphics[width=3.in,clip]{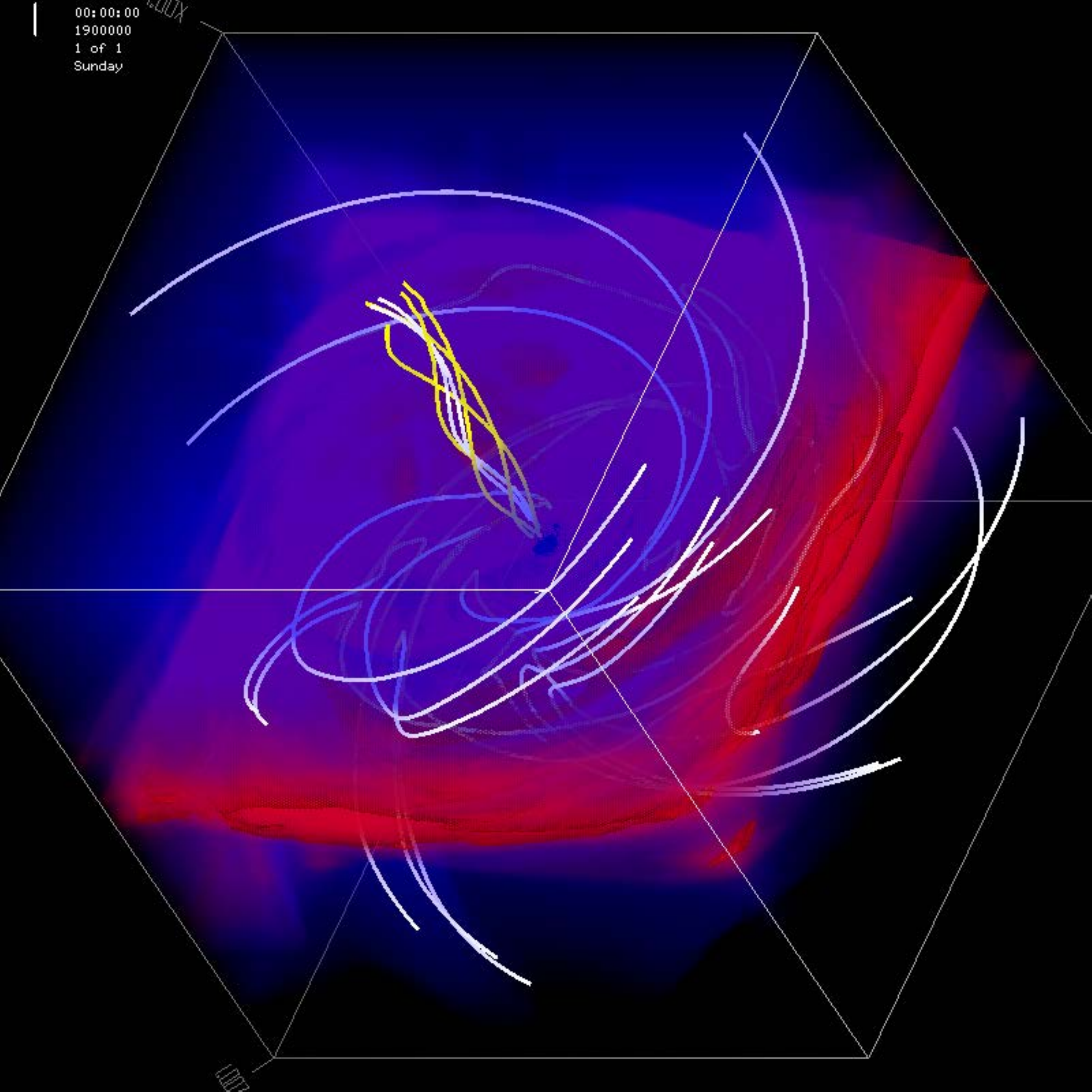}}
\subfigure[Emergence]{\label{fig:b}\includegraphics[width=3.in,clip]{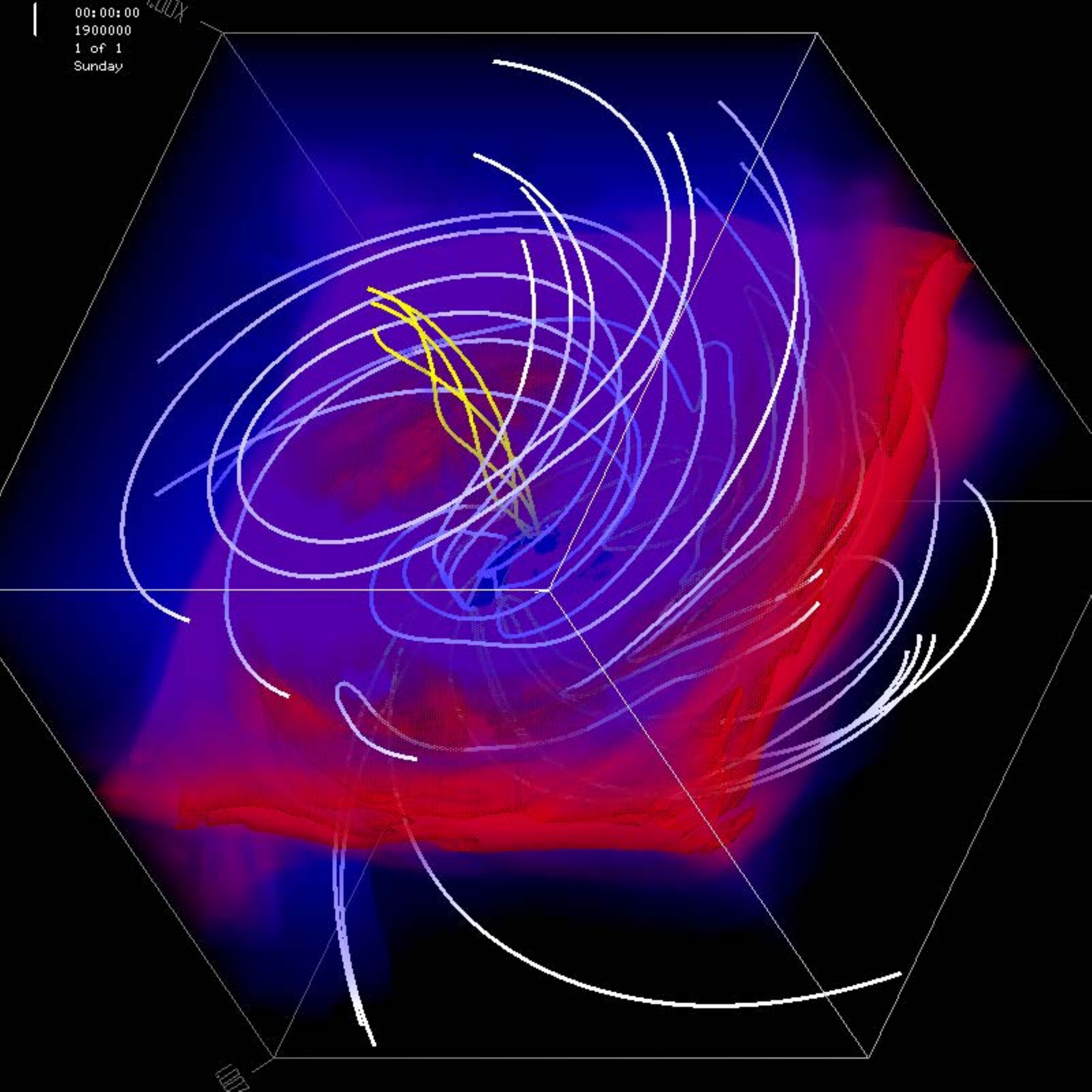}}
\subfigure[Peak]{\label{fig:c}\includegraphics[width=3.in,clip]{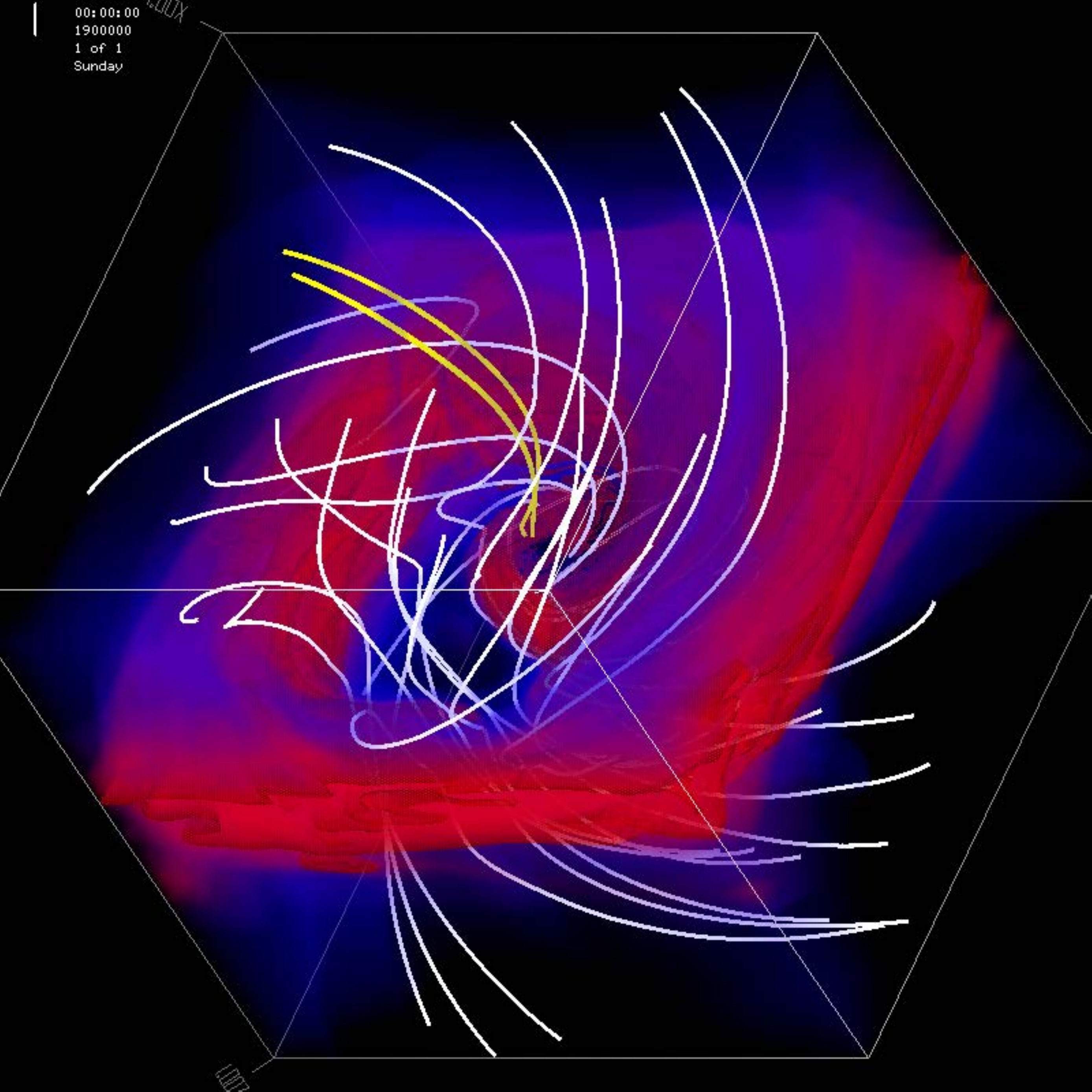}}
\subfigure[Dissipation]{\label{fig:d}\includegraphics[width=3.in,clip]{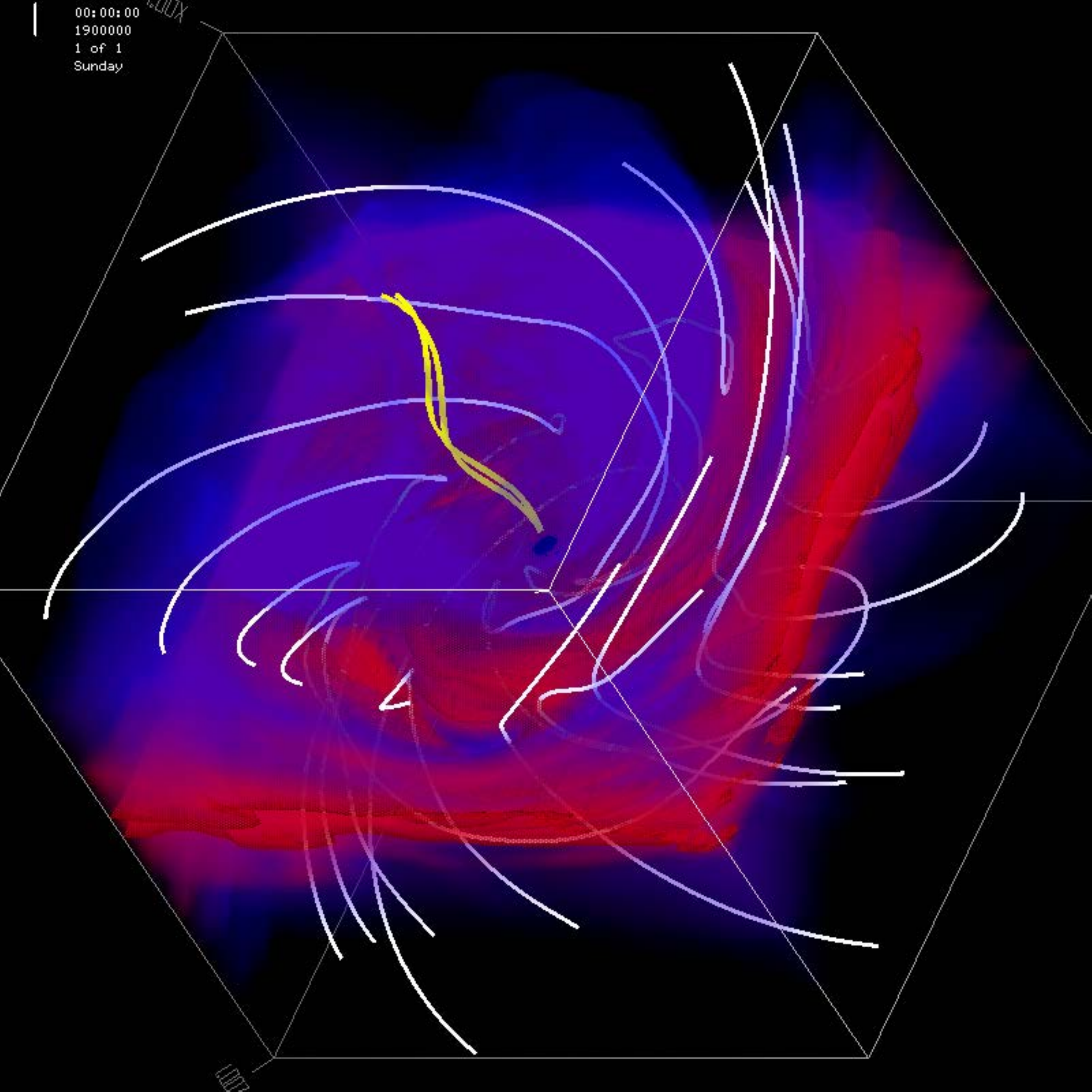}}
\caption{3D renderings of key moments during the lifetime of the RT bubble in a cube $-40r_{g}\leq r\leq 40r_{g}$ on each side. Field lines chosen from the flux probability distribution are shown in white and the field lines fixed to the horizon are yellow. The disk is shown in red, while the corona is blue. From this, we see that as the RT bubble evolves, the magnetic field becomes less cohesive and more turbulent as in Fig.~\ref{fig:c} , then returns to the previous orderly structure as the bubble dissipates (Fig.~\ref{fig:d}. A video of the bubble's evolution can be found here: https://youtu.be/Sfh9O6Nm5Cc and a video of the seedpoints being followed in the midplane can be found here: https://youtu.be/74CuoWN2HjI}
\label{fig:bubblesnaps}
\end{figure*}

\subsection{Stress Analysis}
\label{sec:stress}
Because of the large vertical magnetic field that characterizes the MAD state, we want to consider not only the usual Maxwell stress, 
\begin{equation}
	\alpha_{r\phi} = -\frac{b^r b_\phi}{\langle P_b + P_{gas} \rangle} ,
	\label{eq:radstress}
\end{equation}
but also look at any vertical analog that might lead to angular momentum carried off in a wind, 
\begin{equation}
	\alpha_{z\phi} = -\frac{b^z b_\phi}{\langle P_b + P_{gas} \rangle},
	\label{eq:vertstress}
\end{equation}
where $P_{b}=b^{\mu}b_{\mu}/2$ is the magnetic pressure, $P_{gas}=(\Gamma-1)u_g$ is the ideal gas pressure with adiabatic index $\Gamma=4/3$. $b^{\mu}$ is the contravariant fluid-frame magnetic 4-field and is relate to the laboratory-frame 3-field by $b^{\mu}=B^{\nu}h^{\mu}_{\nu}/u^{t}$, where $h^{\mu}_{\nu}=u^{\mu}u_{\nu}+\delta^{\mu}_{\nu}$ is a projection tensor and $\delta^{\mu}_{\nu}$ is the Kronecker delta function.

Though we know the axisymmetric MRI is suppressed in this simulation as reported in \citet{2016MNRAS.462..636A}, the magnetic RT instability itself creates turbulence.  So to separate the contributions to the angular momentum transport due to turbulence vs. the large vertical field, we decompose the magnetic field into a mean field term plus fluctuations ($b^{\mu}=\langle b^{\mu} \rangle + \delta b^{\mu}$).  Then the (unnormalized) stress decomposition became
\begin{equation}
	\alpha_{\mu\phi} = \langle b^{\mu} b_\phi \rangle + \langle b^{\mu} \rangle \delta b_\phi +\delta b^{\mu} \langle b_\phi \rangle + \delta b^{\mu} \delta b_\phi
    \label{eq:stressdecomp}
\end{equation}

The total radial and vertical stresses as well as the decomposed stress terms are calculated in both the disk and corona. These regions were defined only in terms of angular extent, with no density weighing or other such factors. The angle was chosen using half-height to radius $H/R \approx 0.1$ and the small angle approximation to give $\theta = 0.1$, so the disk is defined as $\frac{\pi}{2}\pm 0.1$ and the corona is $\frac{\pi}{2}\pm 0.1$ to $\frac{\pi}{2}\pm 0.2$. In Fig. \ref{fig:totalstresslong}, the total stress results are plotted from the start of the simulation until after the bubble being studied dissipated (33240 $r_{g}/c$), while in Fig. \ref{fig:radvertlong}, the stress decomposition is shown.

We also examine the differences between the stress in the RT bubble and in the high density area. The regions are separated using the plasma parameter $\beta = P_{gas}/P_b$ as a filter, with $\beta\leq0.1$ considered the RT bubble region and higher values being the higher density region. The total stress in these areas is shown in Fig. \ref{fig:totalstressbubble} and the stress decomposition over the lifetime of the RT bubble is shown in Fig. \ref{fig:radvertbubble}.

\section{Results and Discussion}
\label{sec:results}
\subsection{Visualization}
In the first snapshot Fig.\ref{fig:a}, we see the steady state of the disk. The field on the BH is helical and very tightly wound and the field in the disk has the same general shape, though looser in structure. Moving forward in time to the emergence of the bubble (Fig. \ref{fig:b}), these regions begin to unwind and become less ordered. This is even more prominent in the snapshot of the bubble at its maximal size, Fig. \ref{fig:c}. Here the field is much more tangled, having been swept back by the slower rotation of the bubble in the disk. Also, the field in the jet region is almost vertical, with much less winding. Finally, as the bubble begins to dissipate in Fig.~\ref{fig:d}, the field becomes ordered again with the field in the jet region twisting up once more, as described in \citet{2008ApJ...677..317I}.

\subsection{Stress Analysis}

\begin{figure}
\centering
\includegraphics[width=\columnwidth]{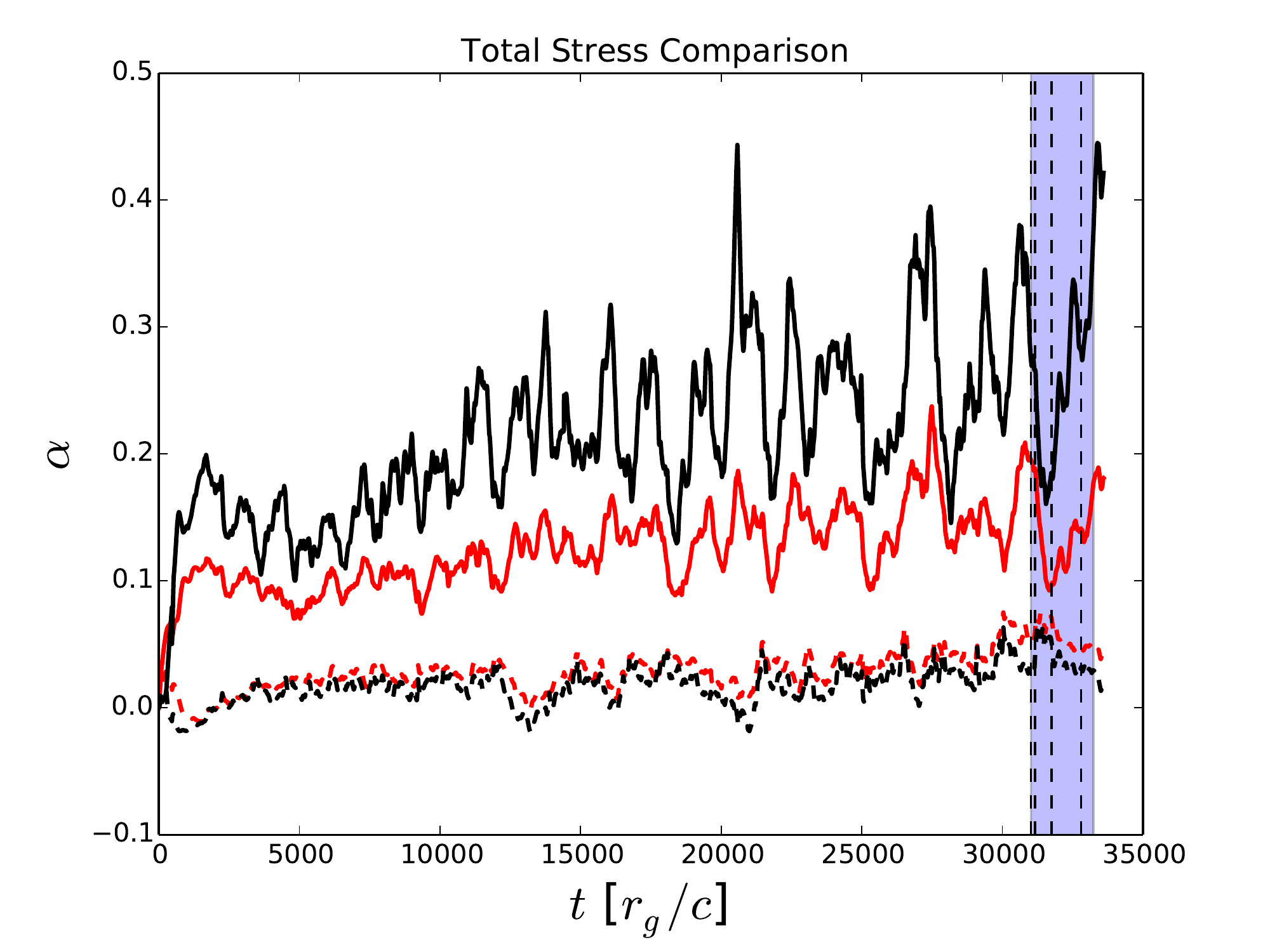}
\caption{Long term stress in the disk (black) and corona (red). The vertical stress is shown as a dashed line, while the radial stress is solid. The time period being studied is highlighted in blue with vertical dashed lines marking the times displayed in Fig. \ref{fig:bubblesnaps}. As the bubble emerges, the radial stress in both regions decreases, but grows as the bubble dissipates, indicating that stress, and therefore accretion, increase after the lifetime of the bubble; however, the vertical stress increases slightly while the bubble is moving through the disk.}
\label{fig:totalstresslong}
\end{figure}

In Fig. \ref{fig:totalstresslong}, we plot the total radial stress integrated over all $\phi$, $10r_{g}\leq r \leq 40r_{g}$, and the $\theta$ ranges given above for the disk and corona are shown. The vertical stress is integrated over the same $\phi$ and radial regions, but the upper and lower halves of the $\theta$ region are calculated separately, then subtracted from each other to ensure that positive stress corresponds to outgoing angular momentum transport. The radial stress through the disk is the dominant term, much higher than the vertical stress even with the large vertical field that characterizes the MAD state. In the corona, radial stress is also higher than the vertical term. However, the vertical stress in the corona is slightly higher than in the disk, indicating the vertical outflows are more prominent there. The radial stress is always positive, corresponding to outward angular momentum transport and enhanced accretion, while the vertical stress is negative at some times, but is usually positive.

\begin{figure}
\centering
\includegraphics[width=\columnwidth]{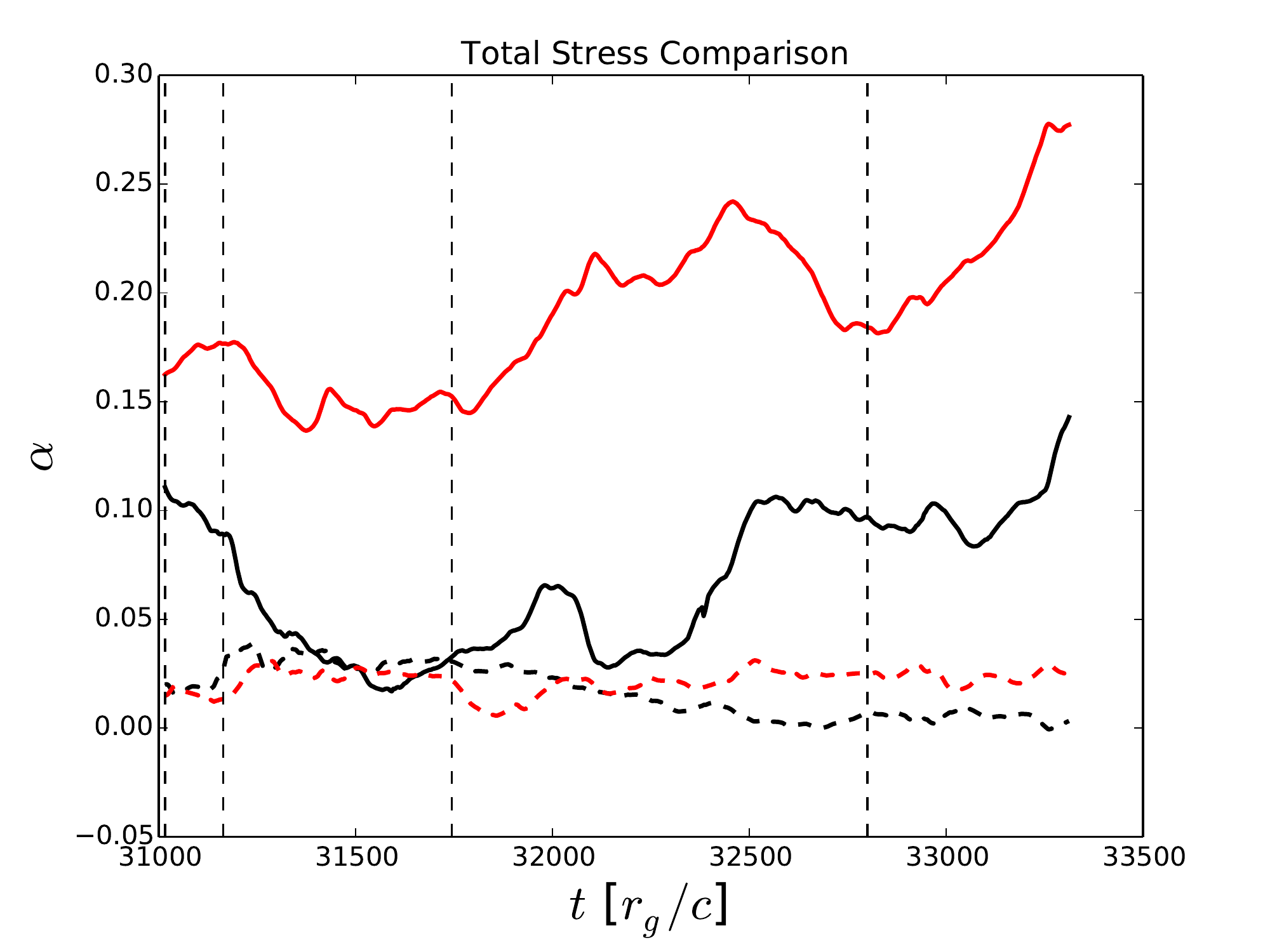}
\caption{Division of stress in the RT bubble (black) and the higher density portion (red) of the disk for the lifetime of the RT bubble. The radial stress is shown as a solid line, while the vertical stress is a dashed line. The vertical dashed lines mark the times displayed in Fig. \ref{fig:bubblesnaps}. As the bubble expands, the radial stress in the bubble decreases, while the vertical stress increases. Outside the bubble, both the radial and vertical stress start to increase as the bubble dissipates around 32000 $r_{g}/c$, with the radial stress almost doubling by the time the bubble disappears.}
\label{fig:totalstressbubble}
\end{figure}

\begin{figure*}
\centering
\subfigure[Radial Disk]{\label{fig:raddisk}\includegraphics[width=3.in,clip]{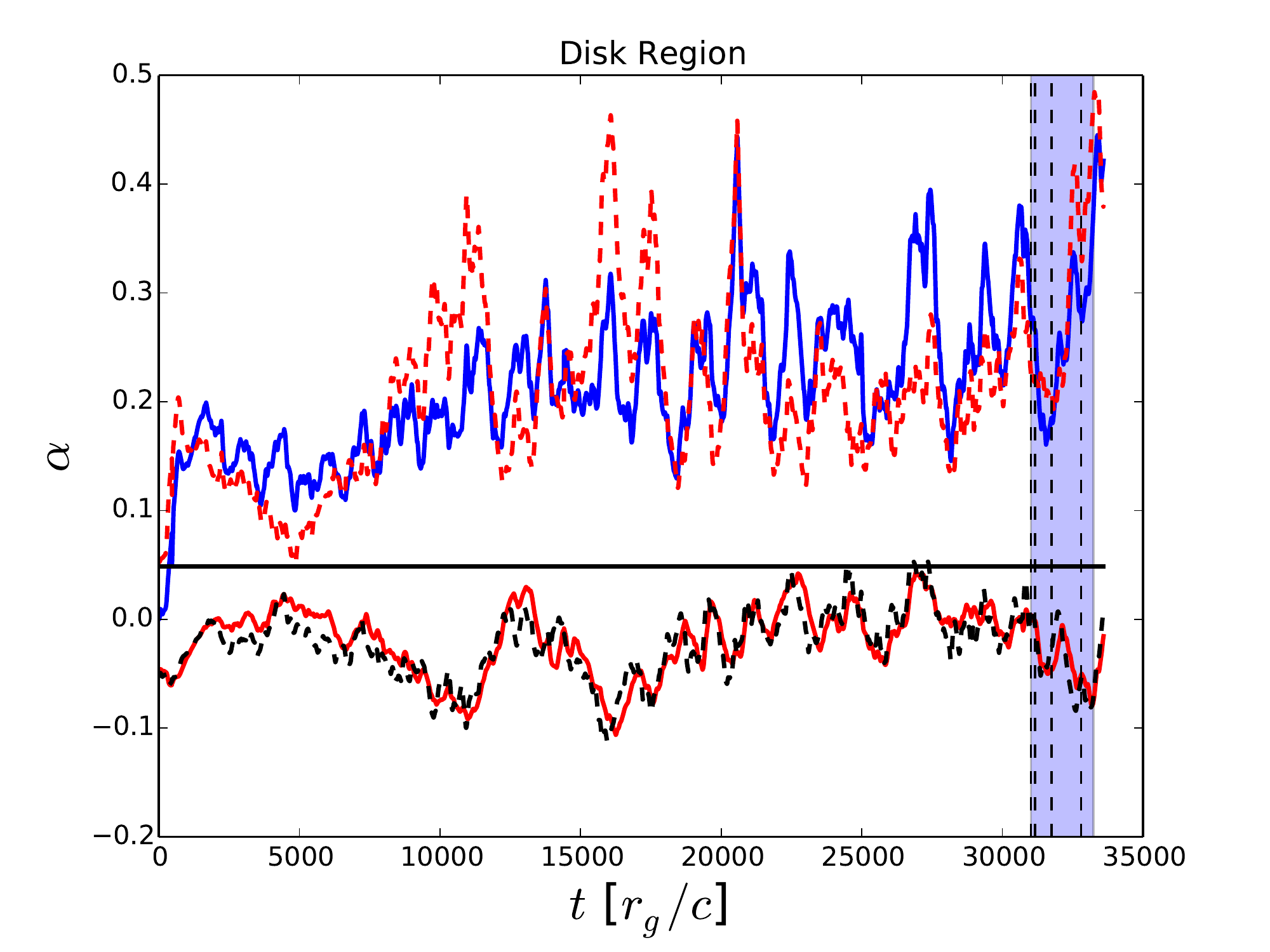}}
\subfigure[Radial Corona]{\label{fig:radcorona}\includegraphics[width=3.in,clip]{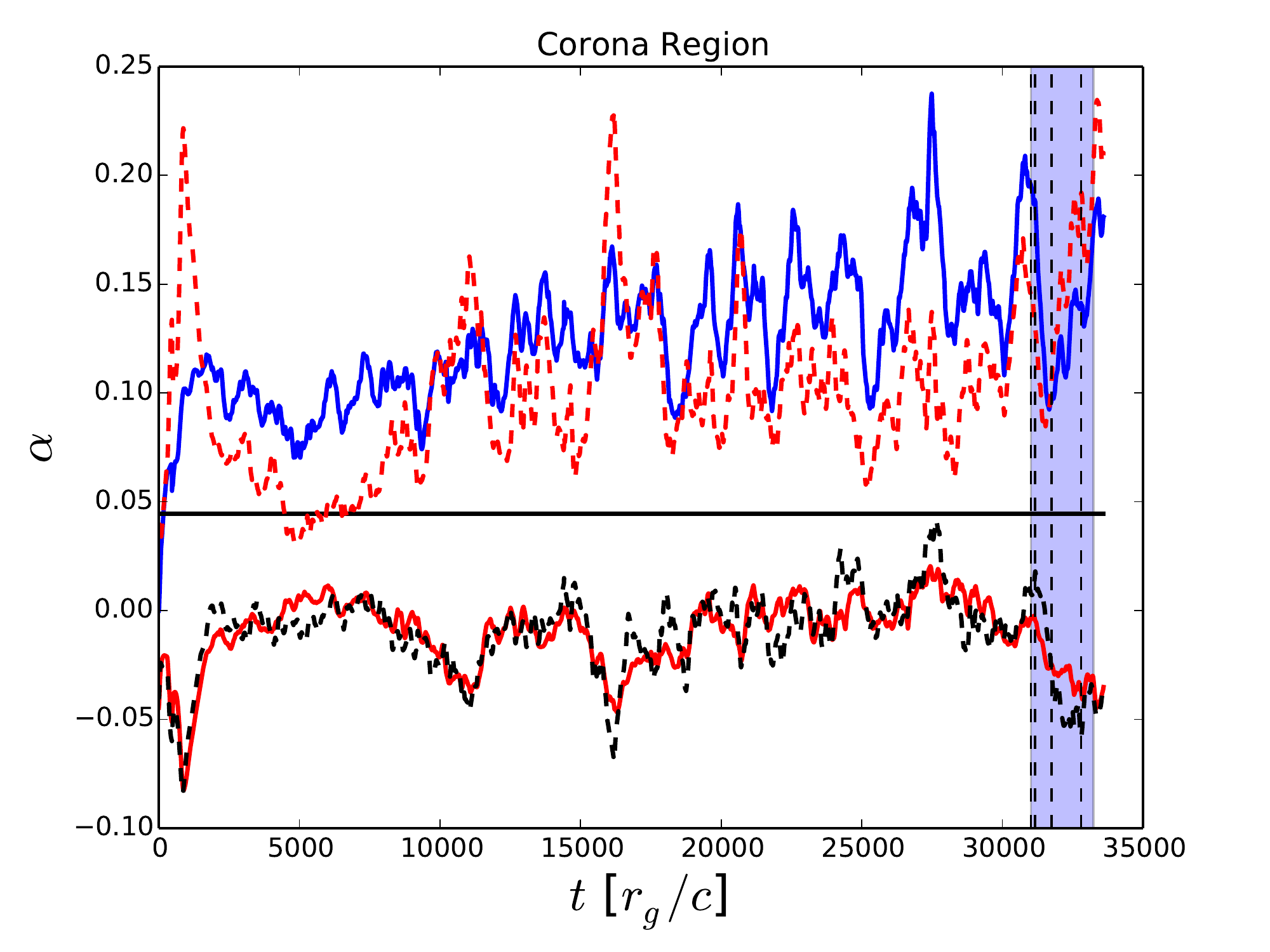}}
\subfigure[Vertical Disk]{\label{fig:vertdisk}\includegraphics[width=3.in,clip]{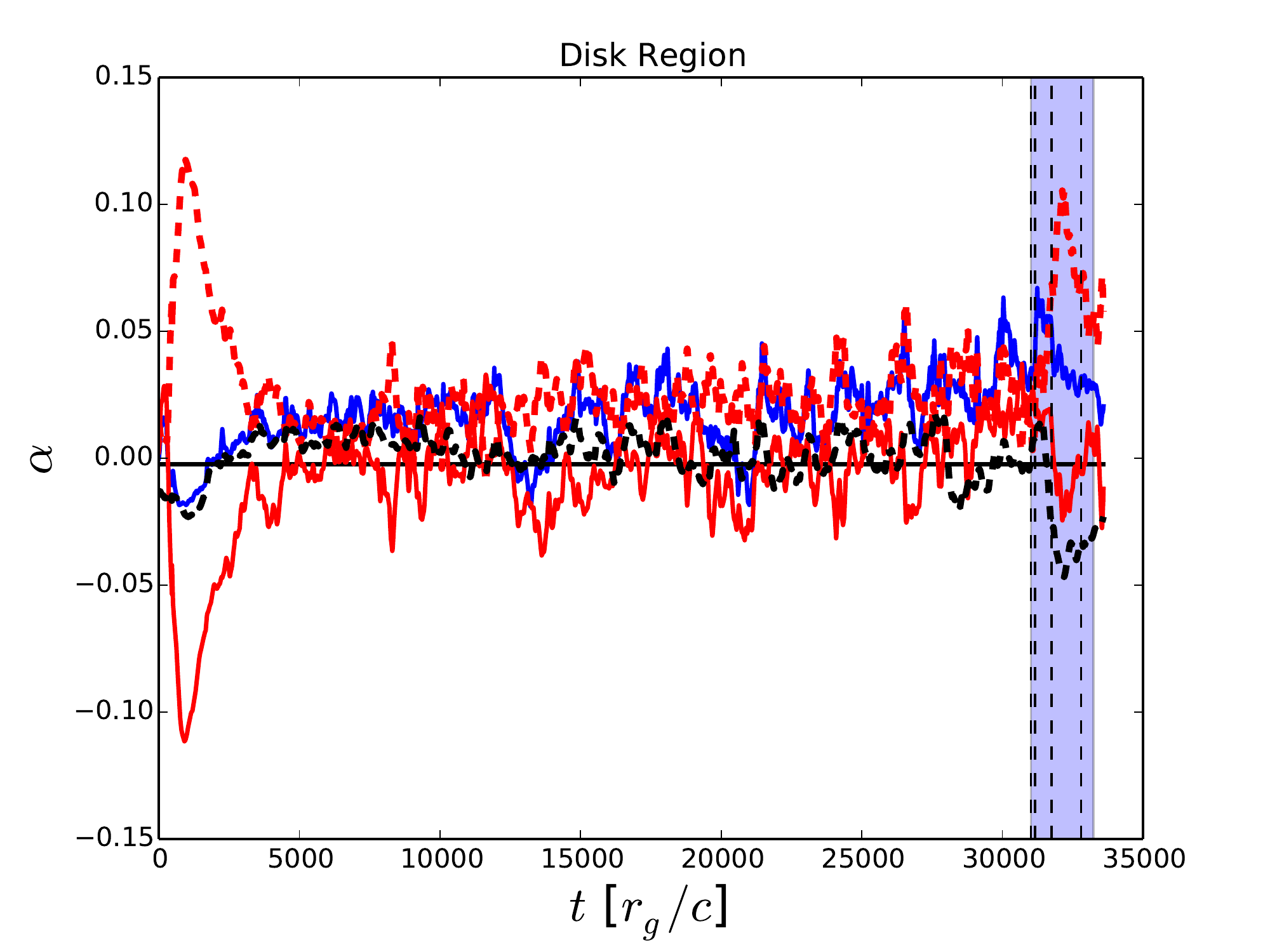}}
\subfigure[Vertical Corona]{\label{fig:vertcorona}\includegraphics[width=3.in,clip]{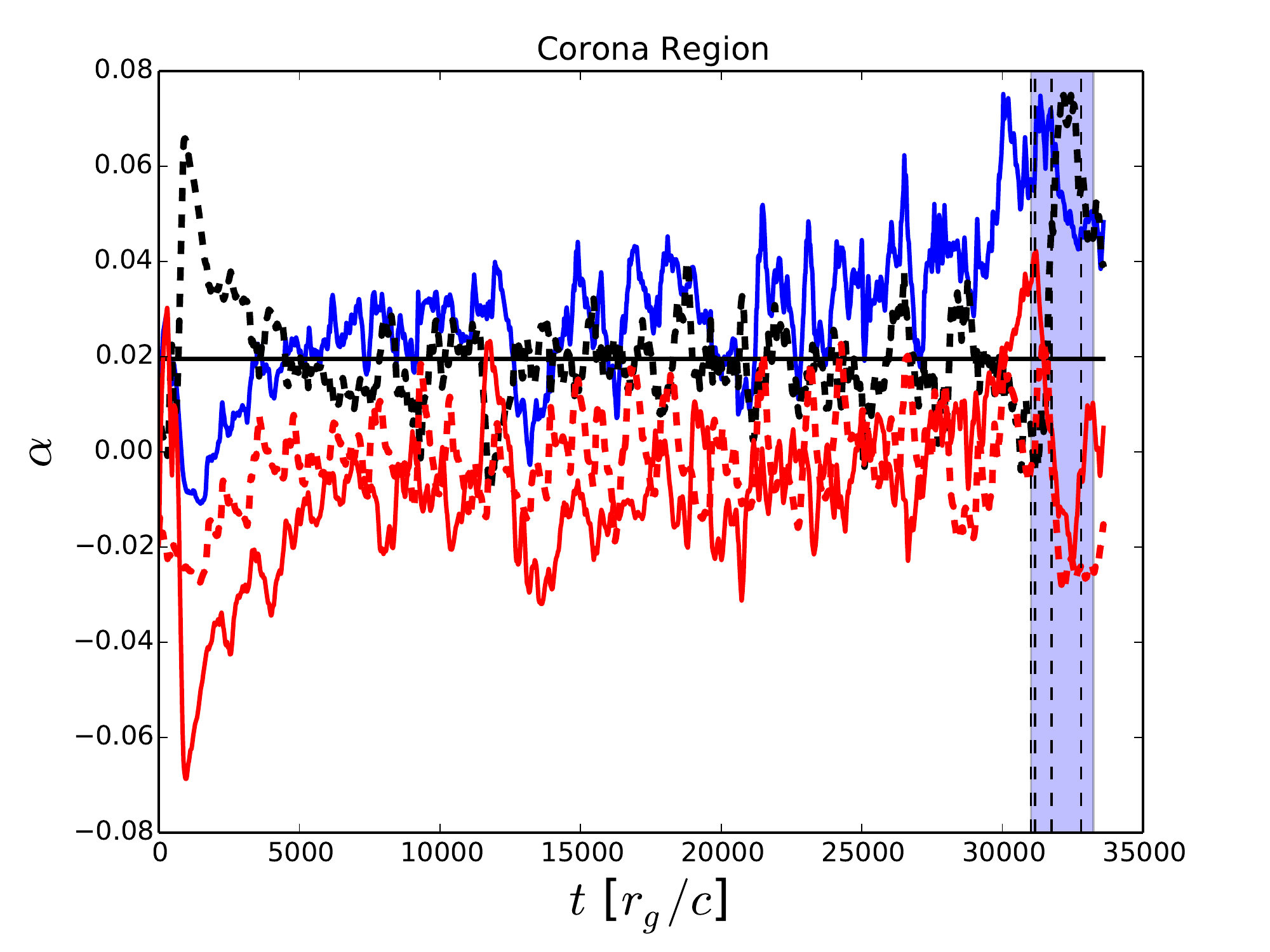}}
\caption{The decomposition of the total stress (blue solid line) into mean field (black solid), purely turbulent (red dashed), and 2 mixed terms ($\langle b^i\rangle \delta b_\phi$ as a red solid lines and $\delta b^i \langle b_\phi \rangle$ as a black dashed line). The time period being studied is highlighted in blue with vertical dashed lines marking the times displayed in Fig. \ref{fig:bubblesnaps}. The upper panels show the radial stress decomposition in the disk (left) and corona (right) while the lower panels are the vertical stress in the disk (left) and corona (right). In both regions, the radial stress is dominated by purely turbulent component, despite the suppression of the MRI. As for the vertical stress, in the disk, the stress components fluctuate around 0, but seem to be net positive. In the corona, the turbulent terms are negative, but the mean field term is positive, meaning it contributes to outward angular momentum transport.}
\label{fig:radvertlong}
\end{figure*}

\begin{figure*}
\centering
\subfigure[Radial stress in the RT bubble]{\label{fig:radbubble}\includegraphics[width=3.in,clip]{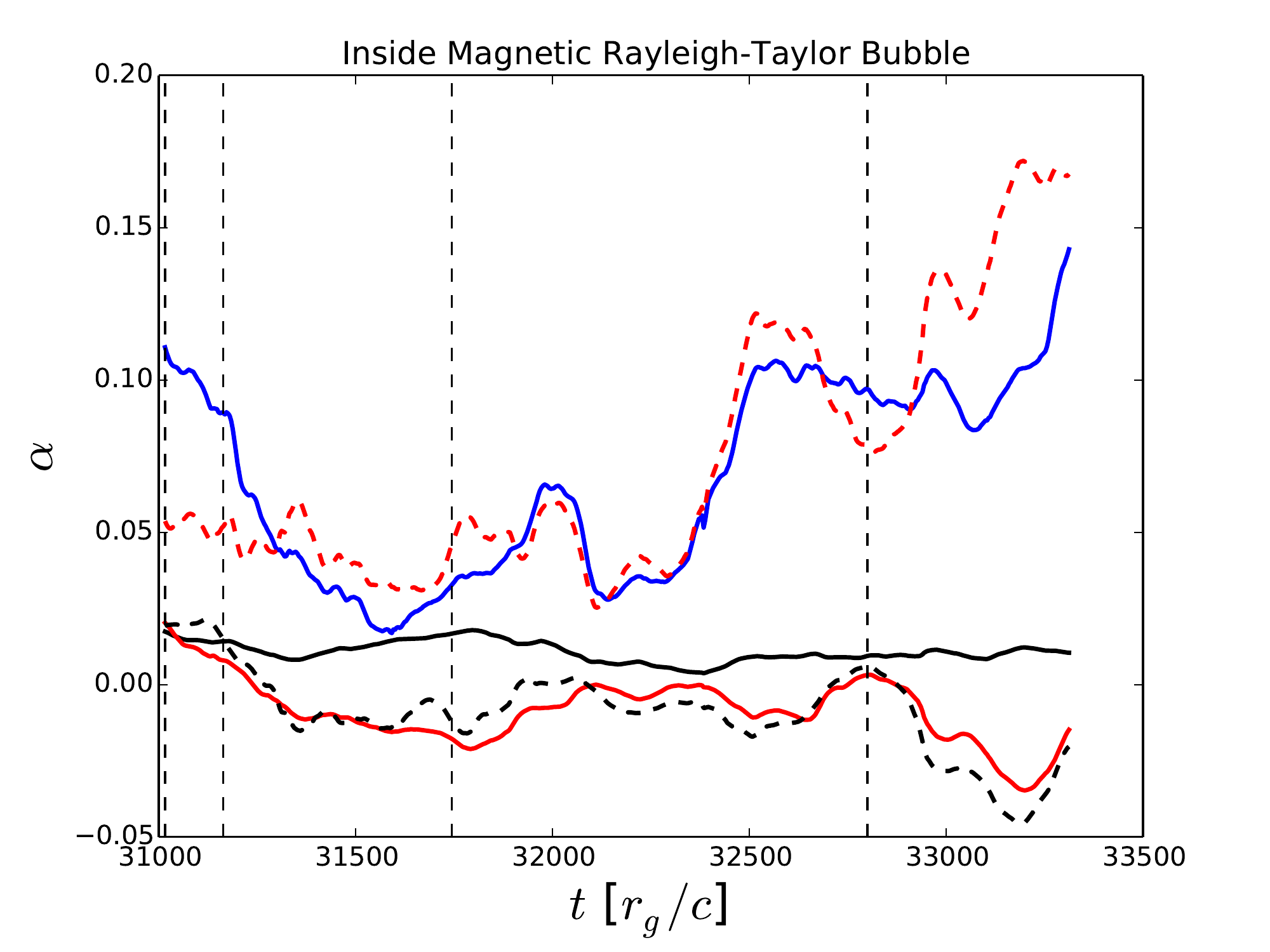}}
\subfigure[Radial stress outside the RT bubble]{\label{fig:radoutbubble}\includegraphics[width=3.in,clip]{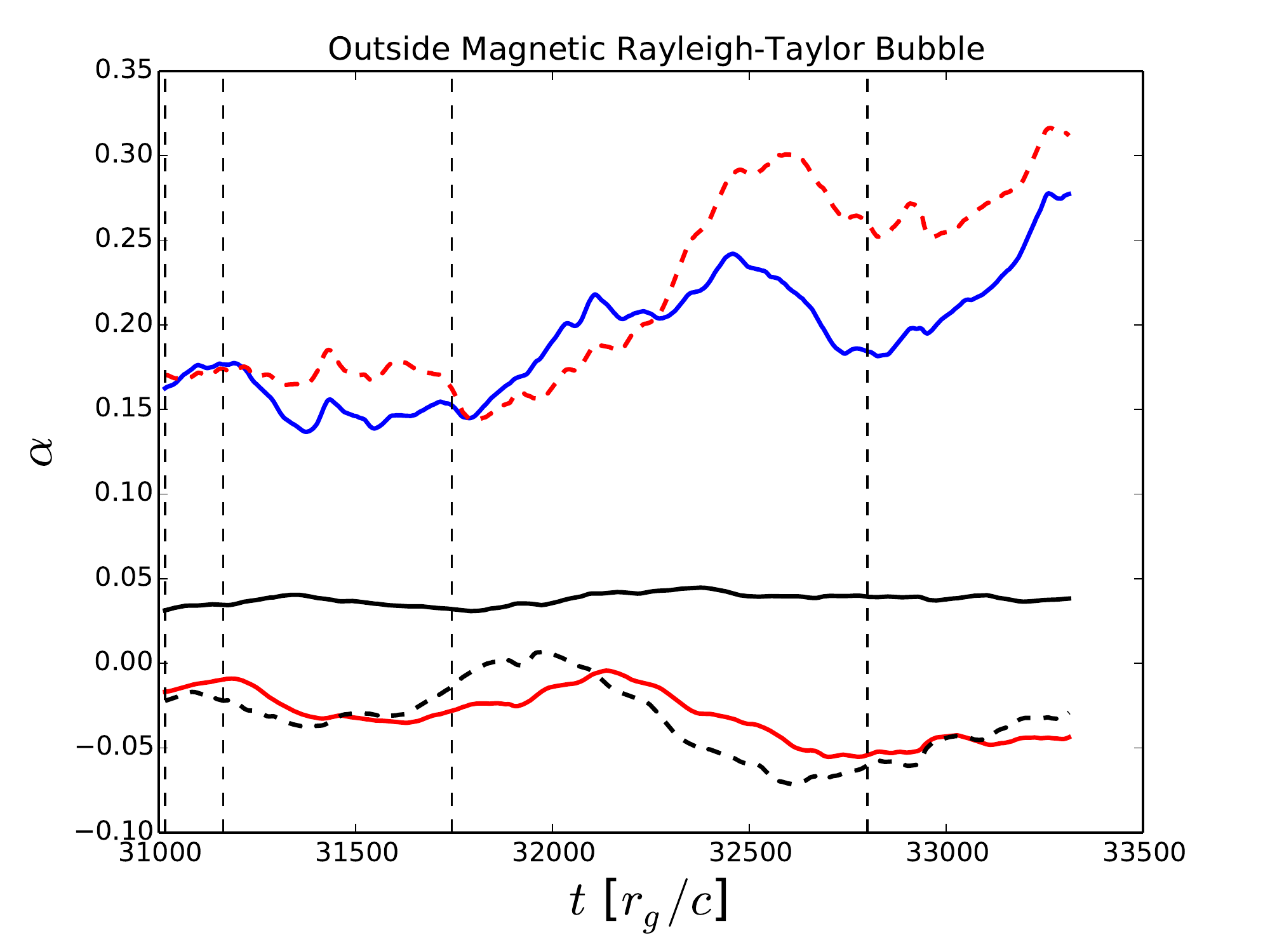}}
\subfigure[Vertical stress in the RT bubble]{\label{fig:vertbubble}\includegraphics[width=3.in,clip]{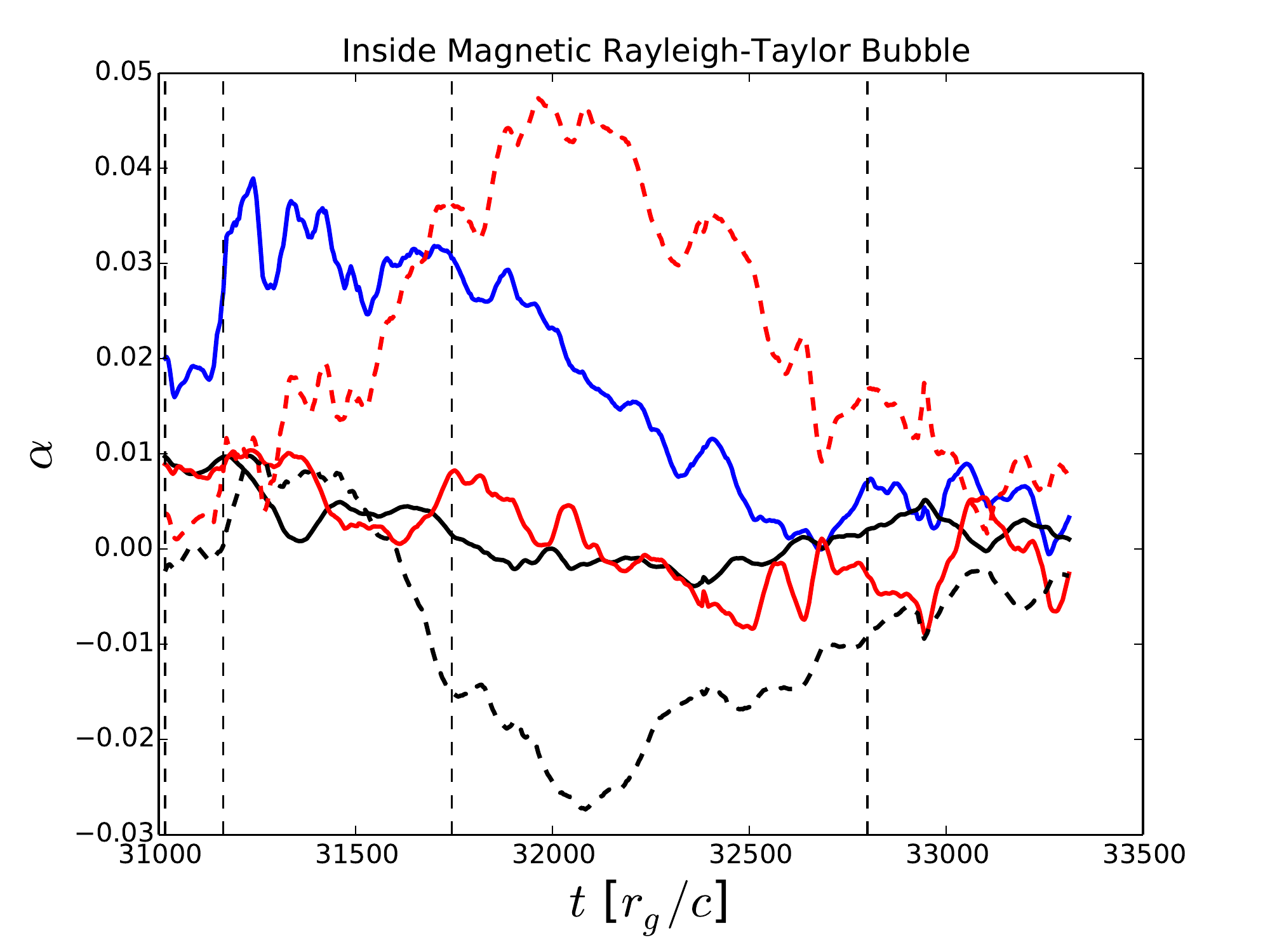}}
\subfigure[Vertical stress outside the RT bubble]{\label{fig:vertoutbubble}\includegraphics[width=3.in,clip]{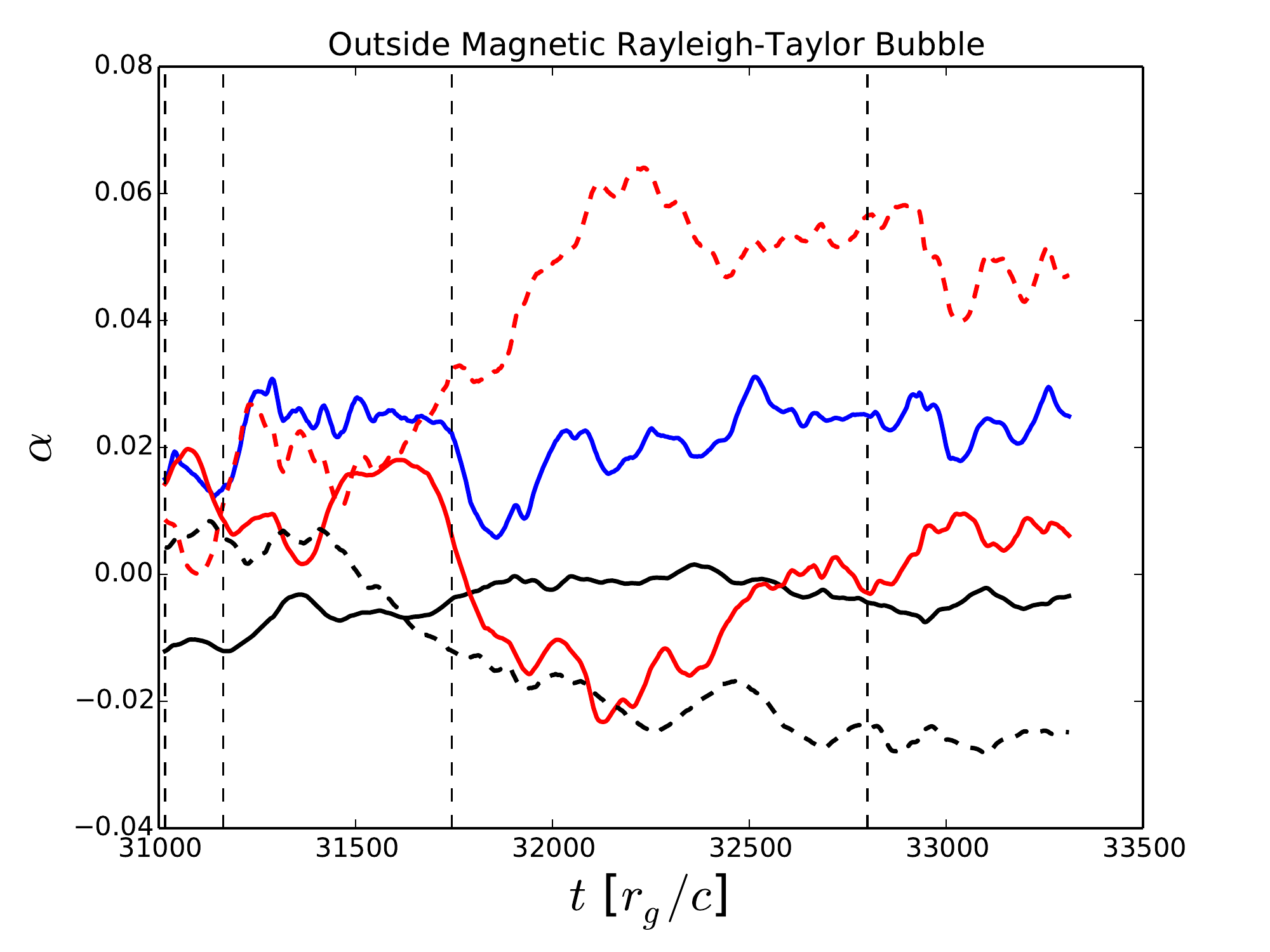}}
\caption{The decomposition of the total stress (blue solid line) into mean field (black solid), purely turbulent (red dashed), and 2 mixed terms ($\langle b^i\rangle \delta b_\phi$ as a red solid lines and $\delta b^i \langle b_\phi \rangle$ as a black dashed line). Vertical dashed lines mark the times displayed in Fig. \ref{fig:bubblesnaps}.  The upper panels show the radial stress decomposition inside (left) and outside (right) the RT bubble while the lower panels are the vertical stress inside (left) and outside (right) the RT bubble. The total vertical stress is close to 0 for the lifetime of the bubble and we see that the component terms mostly balance themselves out. In both regions, the total radial stress is dominated by the purely turbulent term and positive, leading to angular momentum transport.}
\label{fig:radvertbubble}
\end{figure*}

Looking more closely at the terms of the stress decomposition in Fig. \ref{fig:radvertlong}, integrated over the same volume as Fig. \ref{fig:totalstresslong}, we see that the leading component of the radial stress is the purely turbulent term, even though the MRI is suppressed.  This turbulence can easily be driven by the magnetic RT-instability instead of the MRI. This agrees with Fig. \ref{fig:c}, which shows the magnetic field structure becoming less ordered and more turbulent as the RT bubble reaches its maximum extent. The mean field term also contributes positive stress, therefore outward angular momentum transport, in both the disk and corona with roughly the same strength.  The mixed radial stress terms are the smallest components and mostly negative, meaning inward angular momentum transport opposing accretion. In the vertical case, the mean field and the purely turbulent components are roughly equal in magnitude in both regions. In the disk, all terms of the decomposition fluctuate around the mean field term at 0. Other than the $\langle b^{z}\rangle\delta b_{\phi}$ at early times, most of these fluctuations are positive, so angular momentum is transported out of the system except at for these early times. In the corona, though the purely turbulent or mixed terms fluctuate into negative values often, the mean field term is positive, meaning that the ordered field in the corona is prominent in the accretion process. 

At the time the magnetic bubble emerges off the BH, the radial stress decreases while the vertical increases. Overall, this means the stress in the disk is lowered by the emergence of the bubble, hindering accretion. Looking at Fig. \ref{fig:totalstressbubble}, the RT bubble contributes much less radial stress, up to a factor of four less at times. For the total vertical stress, though, the bubble actually has a higher contribution until it starts to dissipate. Once again, the turbulent component is the dominant contribution to the stress, as seen in Fig. \ref{fig:radvertbubble}. The mixed terms again become negative in both parts of the disk. Interestingly, in Fig. \ref{fig:vertbubble}, the mean field term is actually negative until the bubble begins to dissipate and the vertical flux is reprocessed into the disk. As the bubble dissipates, the stress begins to rise again in both parts of the disk and a spike can be seen in Fig. \ref{fig:totalstresslong} after it is reabsorbed by the disk. The strength of the turbulent stress indicates that a secondary instability is creating a turbulent field in the wake of the RT bubble.

\section{Conclusions}
\label{sec:conclusion}
We studied how the magnetic RT instability affects the evolution and angular momentum transport in thin accretion disks in the MAD state by investigating the largest of many RT bubbles produced in the MADiHR simulation from \citet{2016MNRAS.462..636A}. We started by developing the first 3D visualization technique to select and follow magnetic fieldlines in the high flux regions of the bubble.  This showed us how the emergence of the RT bubble leads to less ordered magnetic field in the disk and less twisted up field in the jet region, indicating that the magnetic RT instability is leading to a secondary turbulence in the disk. This visualization method is applicable to many other situations where the magnetic field is disrupted, such as magnetic field inversions or magnetic plasma instabilities in other disk geometries.

We also examined the effects of the RT bubble through stress calculations. We found the dominant contribution to the stress in both the disk and corona is the radial term with the vertical stress up to four times smaller than the radial term. We also saw that the emergence of the RT bubble corresponds to a reduction in radial stress in both the disk and corona, but an increase in the vertical stress.

When the stress terms are decomposed into mean field, purely turbulent, and mixed components, we measure the turbulent component to be dominant, despite the suppression of the MRI. Only in the vertical stress is the mean field contribution as strong as the turbulent one. As seen in the visualization, this turbulence is linked to the emergence of the bubble, so could be due to a secondary instability caused by the magnetic RT instability.

In the bubble itself, the stress is suppressed much more than in the higher density region. Though the vertical stress is much smaller than the radial stress, it is stronger in the bubble than the disk until the bubble starts to dissipate. As seen in the disk, the turbulent field is the dominant driver of the stress, while the mean field term is a factor of 3-4 smaller. Again, this is consistent with the disordered fieldlines seen in the visualization.

These results are limited by the simplified treatment of the thermodynamics of the accretion disk. In the original work done with this simulation, an ad hoc cooling function designed to keep the disk close to the target scale height of $H/R \approx 0.1$, rather than a more complete handling of the thermodynamics.  While this might have kept the disk thinner than a more complex cooling function, we see similar low-density regions created by the magnetic RT instability in simulations of thicker disks, so our results would still hold if the disk in this work were thicker.

The extra dissipation produced by the thin MAD state could lead to more emission from near and within the inner-most stable circular orbit (ISCO).  In prior studies of the emission, the corrections to the spin predictions are minimal except at low spin \citep{2011MNRAS.414.1183K} with higher-energy emission from near the ISCO.  Indeed, the emission is marginally optically thin and could be dominated by non-thermal emission.  Our results suggest that, despite the magnetic field having a large-scale ordered component in the MAD state, the disk is still dominated by turbulent viscosity driven by magnetic RT instabilities and the MRI.  The magnetic RT instabilities form even within the ISCO due to the magnetic field piling-up against the BH, and so they could be an important source of extra emission from within the ISCO.

We hope our visualization and basic analysis of the stress in the accretion disk will drive more analytical and simulation work to understand the origin of angular momentum transport in MADs.  We cannot conclude that the MRI is unimportant in MADs, but the magnetic RT plays an important role in controlling the effective viscosity by developing vigorous turbulence through-out the flow.  While existing analytical analysis of magnetic boundaries in disks cannot be easily applied to these simulations, we plan to next study the magnetic stability directly within the simulation by tracking passive mode growth as done in \citet{2009ApJ...697.1901G}.  By injecting modes, we can directly trace their evolution to see if they behave as expected from the MRI instability or as from the magnetic RT instability.

\section*{Acknowledgements}
This material is based upon work supported by the National Science Foundation under Grant No. NNX14AB46G.




\bibliographystyle{mnras}
\bibliography{biblio2} 






\section{Supporting Information}
Additional Supporting Information may be found in the online version of this article:

movies: A movie of the 3D rendering shown in Fig. \ref{fig:bubblesnaps}, showing the time evolution of the fieldlines in the disk (white) and jet regions (yellow) is found at https://youtu.be/Sfh9O6Nm5Cc. A 2D movie of these seedpoints plotted over the probability function described in Eqn. \ref{eqn: prob} calculated in the disk midplane is found at https://youtu.be/74CuoWN2HjI.


\bsp	
\label{lastpage}
\end{document}